\documentclass[3p,times]{elsarticle}

\usepackage{ecrc}

\volume{00}

\firstpage{1}

\journalname{Nuclear Physics A}

\runauth{E.~Iancu and D.N.~Triantafyllopoulos}

\jid{npa}

\jnltitlelogo{Nuclear Physics A}

\usepackage{amssymb}

\biboptions{square,comma,numbers,sort&compress}

\usepackage[figuresright]{rotating}
\usepackage{bm,amsmath,amssymb}

\long\def\comment#1{ }
\newcommand{\eqn}[1]{Eq.~(\ref{#1})}
\newcommand{\beq}{\vspace*{-0.1cm}\begin{equation}}
\newcommand{\eeq}{\vspace*{-0.1cm}\end{equation}}

\newcommand{\dif}{{\rm d}}
\newcommand{\rmd}{{\rm d}}
\newcommand{\rme}{{\rm e}}
\newcommand{\rmi}{{\rm i}}
\newcommand{\rmP}{{\rm P}}
\newcommand{\rmtr}{{\rm tr}}

\newcommand{\mcal}{\mathcal}

\newcommand{\bk}{\bm{k}}

\newcommand{\bp}{\bm{p}}
\newcommand{\bx}{\bm{x}}
\newcommand{\by}{\bm{y}}
\newcommand{\bu}{\bm{u}}

\newcommand{\bz}{\bm{z}}

\newcommand{\br}{\bm{r}}
\newcommand{\bbx}{\bm{\bar{x}}}
\newcommand{\bby}{\bm{\bar{y}}}
\newcommand{\bbu}{\bm{\bar{u}}}

\newcommand{\pd}{{\phantom{\dagger}}}

\begin{document}

\begin{frontmatter}



\dochead{}

\title{JIMWLK evolution: from color charges to rapidity correlations}


\author[ei]{E.~Iancu}
\author[dnt]{D.N.~Triantafyllopoulos\corref{cor1}}
\address[ei]{Institut de Physique Th\'{e}orique de Saclay, F-91191 Gif-sur-Yvette, France}
\address[dnt]{ECT* and Fondazione Bruno Kessler, Strada delle Tabarelle 286, I-38123 Villazzano (TN), Italy}

\begin{abstract}
We study multi-particle production with rapidity correlations in high--energy p+A collisions. In the context of the Color Glass Condensate, the evolution for such correlations is governed by a generalization of the JIMWLK equation which evolves the strong nuclear fields both in the amplitude and in the complex conjugate one. We give the equivalent Langevin formulation, whose main ingredient is the color charge density linked to a projectile parton (a Wilson line).

\end{abstract}


\end{frontmatter}


Multi--particle correlations in hadronic collisions at RHIC and the LHC, and in particular long--range ones in pseudo-rapidity $\Delta \eta$, provide information about phenomena related to high--parton densities. Causality suggests that such correlations are built at early times and thus may contain data about the incoming hadronic wave functions, but they may be affected by final--state interactions and  collective phenomena. For example, the `ridge' in A+A collisions seems to be a combination of initial--state correlations in rapidity and final--state collective flow leading to azimuthal collimation. But such an interpretation has been questioned by the discovery of similar phenomena in p+A or even p+p collisions in events with high multiplicity, where strong final--state effects were a priori not expected. A difficulty in studying all such initial--state correlations is the lack of factorization for calculating multi--particle production in the presence of multiple scattering. For p+A collisions, we proposed a solution to this problem \cite{Iancu:2013uva} by constructing a suitable Langevin equation. (See \cite{JalilianMarian:2004da,Hentschinski:2005er,Kovner:2006ge,Kovner:2006wr} for previous related work.)
     
\begin{figure}
\begin{center}
\includegraphics[width=0.23\textwidth,angle=0]{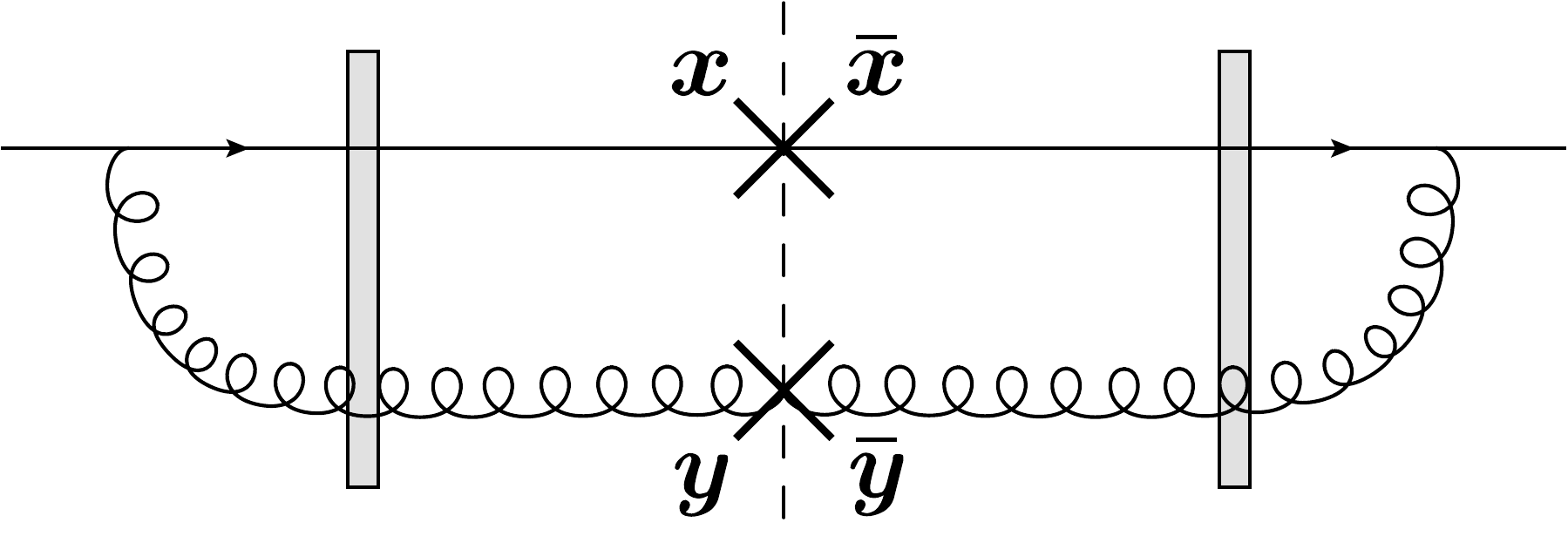} \quad
\includegraphics[width=0.23\textwidth,angle=0]{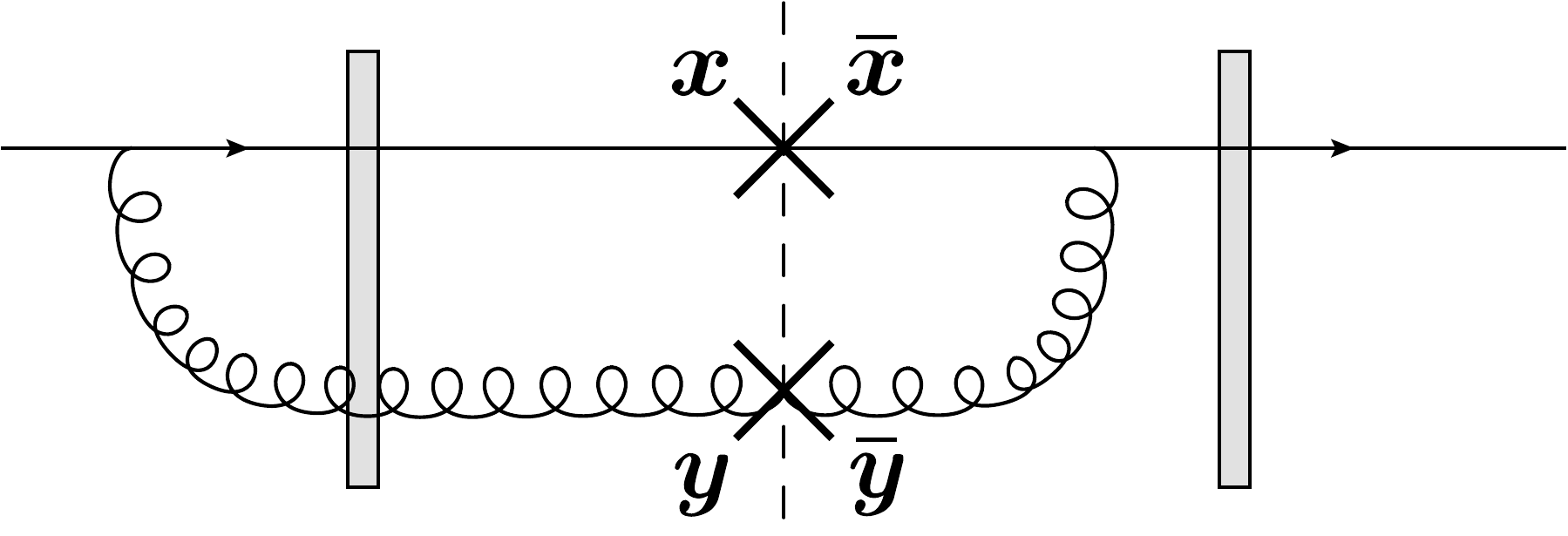} \quad
\includegraphics[width=0.23\textwidth,angle=0]{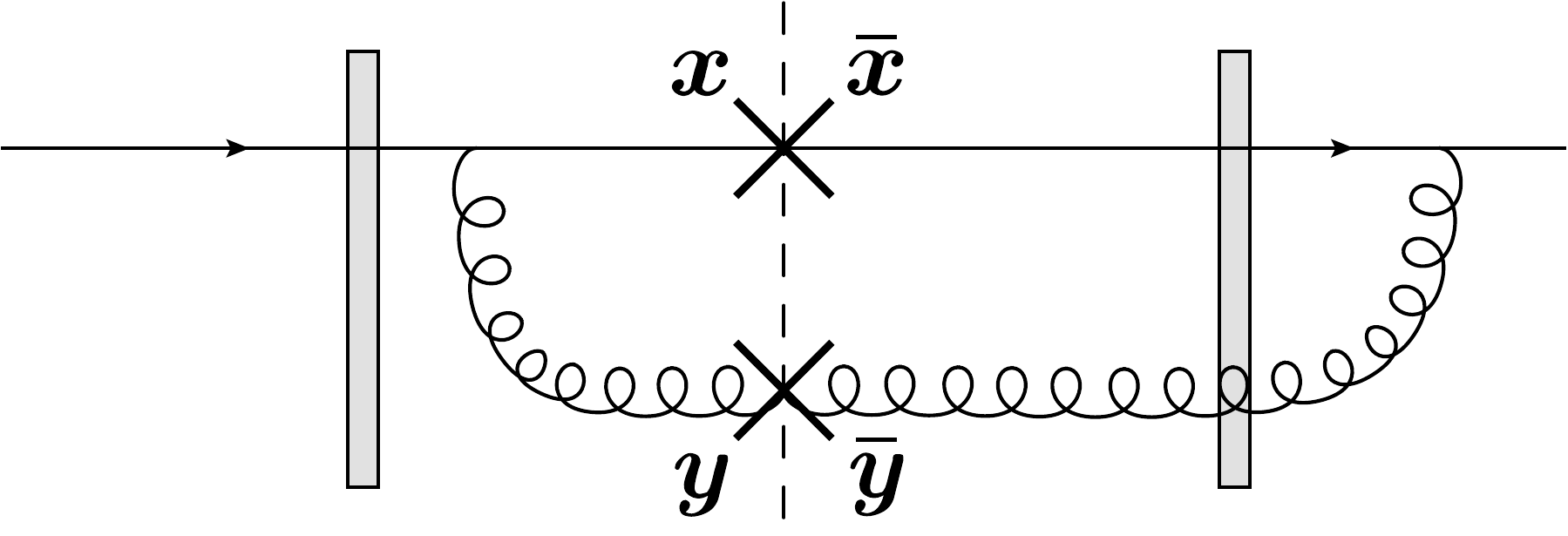} \quad
\includegraphics[width=0.23\textwidth,angle=0]{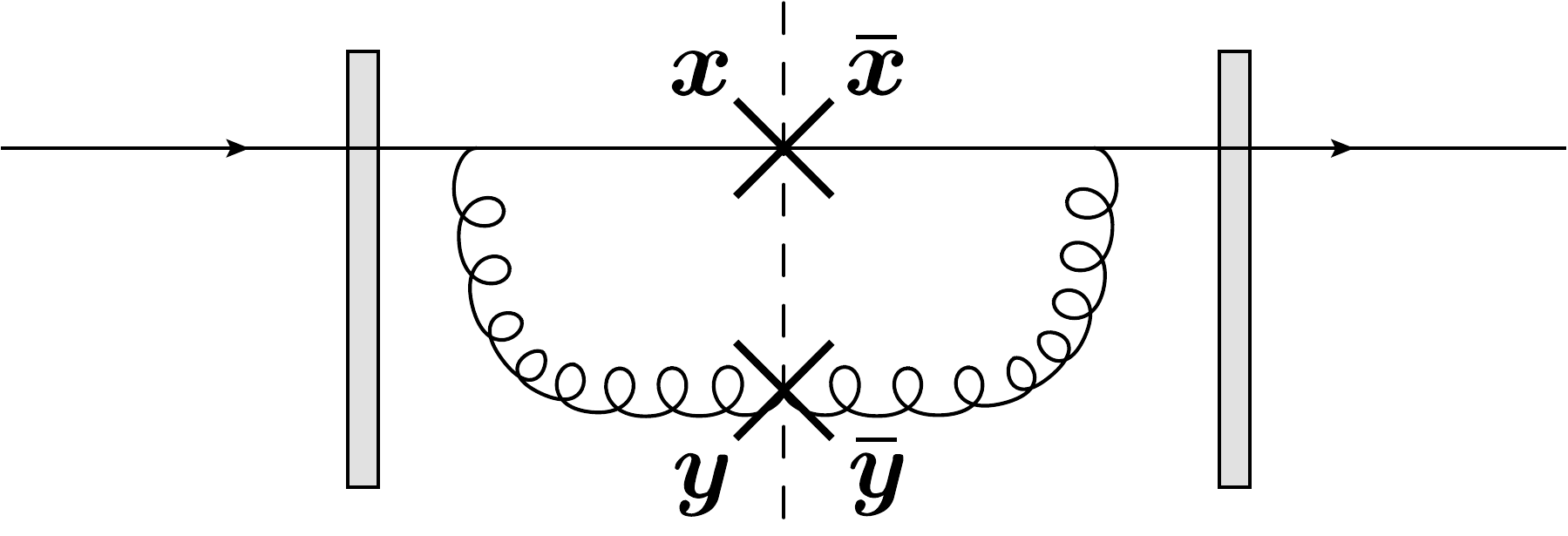}
\end{center}
\vspace*{-0.5cm}
\caption{\label{fig:qgprodsame}The four diagrams for the production of a quark and a gluon at the same rapidity. A cross stands for each parton produced.}
\vspace*{-0.2cm}
\end{figure}

Consider quark-gluon production in the fragmentation region of the proton in p+A collisions. A quark from the proton with a large longitudinal momentum fraction scatters off the nucleus (a shockwave by Lorentz contraction) and emits a gluon, either before or after the scattering.  The diagrams for the product of the direct amplitude (DA)  and the complex conjugate one (CCA) are shown in Fig.~\ref{fig:qgprodsame}. If the gluon is much softer than its parent quark, the cross-section can be computed by acting with the {\em soft gluon production Hamiltonian} on the {\em quark generating functional} \cite{Iancu:2013uva}~:
 \beq
 \label{samey}
 \frac{\dif \sigma^{pA\to qgX}}{\dif Y\,
 \dif^2 \bp \,  \dif^2 \bk} = xq(x)\,
 \frac{1}{(2\pi)^4} 
 \int \dif^2\bx\,\dif^2\bbx\, 
 \rme^{-\rmi \bp \cdot (\bx - \bbx)} 
 \big\langle H_{\rm prod}(\bk)\, 
 {S}^<_{\bx\bbx}   \big|_{\bar{V}=V}
 \big\rangle_Y.
 \eeq
Here $\bp$ and $\bk$ are the transverse momenta of the quark and the gluon, $Y$ is their common rapidity w.r.t. the valence d.o.f.~of the target, $xq(x)$ is the collinear quark p.d.f.~in the proton, and the other notations will be shortly explained.

To understand the {\em generating functional}, consider first quark--production, in which a large--$x$ quark from the proton scatters multiply off the nucleus and acquires a transverse momentum $\bp$. The single-inclusive yield is given by
 \beq
 \label{ptbroad}
 \frac{\rmd N}{\rmd^2\bp} 
 \,=\,xq(x)\,
 \frac{1}{(2\pi)^2} 
 \int \dif^2\br\,
 \rme^{-\rmi \bp \cdot \br} 
 \langle {S}_{\bx\bbx}  \rangle_Y\,,
 \eeq
with $\br\equiv \bx - \bbx$. In the above $\langle {S}_{\bx\bbx}  \rangle_Y$ is the $S$--matrix for a fictitious quark--antiquark dipole
scattering off the nucleus, in which the quark leg at $\bx$ is the physical quark in the DA, 
while the antiquark leg at $\bbx$ is the physical quark in the CCA. The charge of each fermion undergoes color precession in the target field and if the projectile is a right-mover with light-cone time $x^+$, the $S$--matrix {\em operator} (corresponding to a given configuration of the target field $A^-$) reads
 \beq
 \label{Sdip} 
 {S}_{\bx\bbx} [V]  
 \equiv  (1/N_c)\,
 \rmtr\big\{ V^{\dagger}_{\bx}{V}_{\bbx} \big\}
 \qquad \mbox{with} \qquad
 V^{\dagger}_{\bx} = 
 \rmP \exp\left [\rmi g \int  
 \dif x^+ \,A^-_a(x^+, {\bx}) t^a \right]\,,
 \eeq
where  $V^{\dagger}_{\bx}$ and ${V}_{\bbx}$  are Wilson lines describing the color precession
in the DA and respectively the CCA. The physical $S$--matrix follows after averaging  over all the configurations of $A^-$ with the CGC weight function $W_Y[A^-]$ \cite{Gelis:2010nm}~:
 \beq
 \label{save}
 \big\langle S_{\bx\by}\big\rangle_Y = 
 \int \mcal{D}A^- \, W_Y[A^-]\,
 (1/N_c)\,
 \rmtr \big\{ V^{\dagger}_{\bx}{V}_{\by} \big\}.
 \eeq
The color fields $A^-$ which matter for this process represent small-$x$ gluons, i.e.~gluons close to the rapidity of the produced quark and hence are widely separated in rapidity from the valence d.o.f.~the nucleus. The CGC weight function $W_Y[A^-]$ encodes this nonlinear (due to parton saturation) evolution as given by the JIMWLK equation \cite{Gelis:2010nm}. 

The {\em quark generating functional} ${S}^<_{\bx\bbx} $ is a generalization of the
$S$--matrix operator in \eqn{Sdip} to cases where one needs to distinguish between the Wilson lines
in the DA ($V^{\dagger}_{\bx}$, ${V}_{\by}$) and respectively the CCA
($\bar{V}^{\dagger}_{\bbx}$, $\bar{V}_{\bby}$),  and reads \cite{Iancu:2013uva}
 \beq
 \label{s12}
 S^<_{\bx\bbx} [V, \bar{V}]=
 (1/N_c)\, \rmtr \big\{V^{\dagger}_{\bx} 
 \bar{V}_{\bbx}\big\}\,.
 \eeq
This is a functional of the Wilson lines $V$ and $\bar{V}$ (or $A^-$ and $\bar{A}^-$), which must be treated as independent functions at intermediate stages in the calculations. It is only after `emitting the gluons'
by acting with $H_{\rm prod}(\bk)$, that one has to identify $V$ and $\bar{V}$ with
each other and with the {\em physical} target field, to be eventually averaged out according to \eqn{save}.

The {\em production Hamiltonian} $H_{\rm prod}(\bk)$ is an operator which describes the emission of a soft gluon from color sources (fast partons) represented as Wilson lines within the generating functional. It reads \cite{Kovner:2006ge,Kovner:2006wr}
 \beq
 \label{hprod}
 H_{\rm prod}(\bk) = \frac{1}{4\pi^3}\,
 \int \dif^2 \by\, \dif^2 \bby\,
 \rme^{-\rmi \bk \cdot (\by - \bby)}
 \int \dif^2 \bu\,\dif^2 \bbu\,
 \mcal{K}^i_{\by\bu} \, \mcal{K}^i_{\bby\bbu}
 \big[L^a_{\bu} - U^{\dagger ab}_{\by} 
 R^b_{\bu}\big]
 \big[\bar{L}^a_{\bbu} - 
 \bar{U}^{\dagger ac}_{\bby} 
 \bar{R}^c_{\bbu}\big]\,,\qquad
 \mathcal{K}^i_{\by\bu} \equiv 
 \frac{(\by - \bu)^i}{(\by - \bu)^2}\,.
 \eeq
The operators $R^a_{\bu}$ and $L^a_{\bu}$ (and in the CCA $\bar{R}^a_{\bbu}$ and $\bar{L}^a_{\bbu}$) generate soft gluon emissions before and after the scattering when acting on the Wilson lines.The adjoint Wilson lines $U^{\dagger}_{\by}$ and $\bar{U}^{\dagger}_{\bby}$ stand for the emitted gluon in the DA and the CCA. $\mcal{K}^i_{\by\bu}$ is the propagator of the emitted 
gluon in the transverse plane, aka the  Weizs\"{a}cker-Williams kernel. The Fourier transform from $\by-\bby$ to $\bk$ ascribes a transverse momentum to the produced gluon. When acting on ${S}^<_{\bx\bbx} $, the production Hamiltonian generates the diagrams shown in Fig.~\ref{fig:qgprodsame}. $R^a_{\bu}$ and $L^a_{\bu}$ are Lie derivatives which act on the Wilson line $V^{\dagger}_{\bx}$ as infinitesimal gauge rotations to the right and to the left
(and similarly for the action of $\bar{R}^a_{\bbu}$ and $\bar{L}^a_{\bbu}$ on $\bar{V}^{\dagger}_{\bbx}$)~:
 \beq
 \label{rl}
 R^a_{\bu} V^{\dagger}_{\bx} = 
 \rmi g \delta_{\bu\bx} V^{\dagger}_{\bx} t^a,
 \qquad
 L^a_{\bu} V^{\dagger}_{\bx} = 
 \rmi g \delta_{\bu\bx} t^a V^{\dagger}_{\bx} 
 = U^{\dagger ab}_{\bu}\, 
 R^b_{\bu} V^{\dagger}_{\bx}.
 \eeq  
This makes it clear that the following operators represent the color charge density (more precisely,
the density of the `plus' component of the color current) at $\bu$ associated
with a quark at $\bx$ {\em before} and respectively {\em after} the scattering:  
 \beq
 \label{chargerl}
 \mcal{R}^a_{\bu\bx} \equiv  - \rmi V_{\bx}
 R^a_{\bu} V^{\dagger}_{\bx} = 
 g \delta_{\bu\bx} t^a,
 \qquad
 \mcal{L}^a_{\bu\bx} \equiv -\rmi V_{\bx}
 L^a_{\bu} V^{\dagger}_{\bx} = 
 g \delta_{\bu\bx} U^{\dagger ab}_{\bu} t^b.
 \eeq  
Before the scattering the charge is independent of the target, but after the scattering it gets rotated to $\mcal{L}^a_{\bu\bx}= U^{\dagger ab}_{\bu} \mcal{R}^a_{\bu\bx}$.

When the rapidity difference $\Delta Y$
between the `fast' quark and the `soft' gluon is relatively large, $\Delta Y \gtrsim 1/\alpha_s$, one has to take into account the effects of the high-energy evolution within $\Delta Y$,
i.e., the emission of unresolved gluons at intermediate rapidities, between the two
measured particles. By appropriately choosing
the frame, one can associate these emissions with either the projectile, or the 
nucleus, and it is instructive to consider both these points of view. 

\begin{figure}
\begin{center}
\includegraphics[width=0.35\textwidth,angle=0]{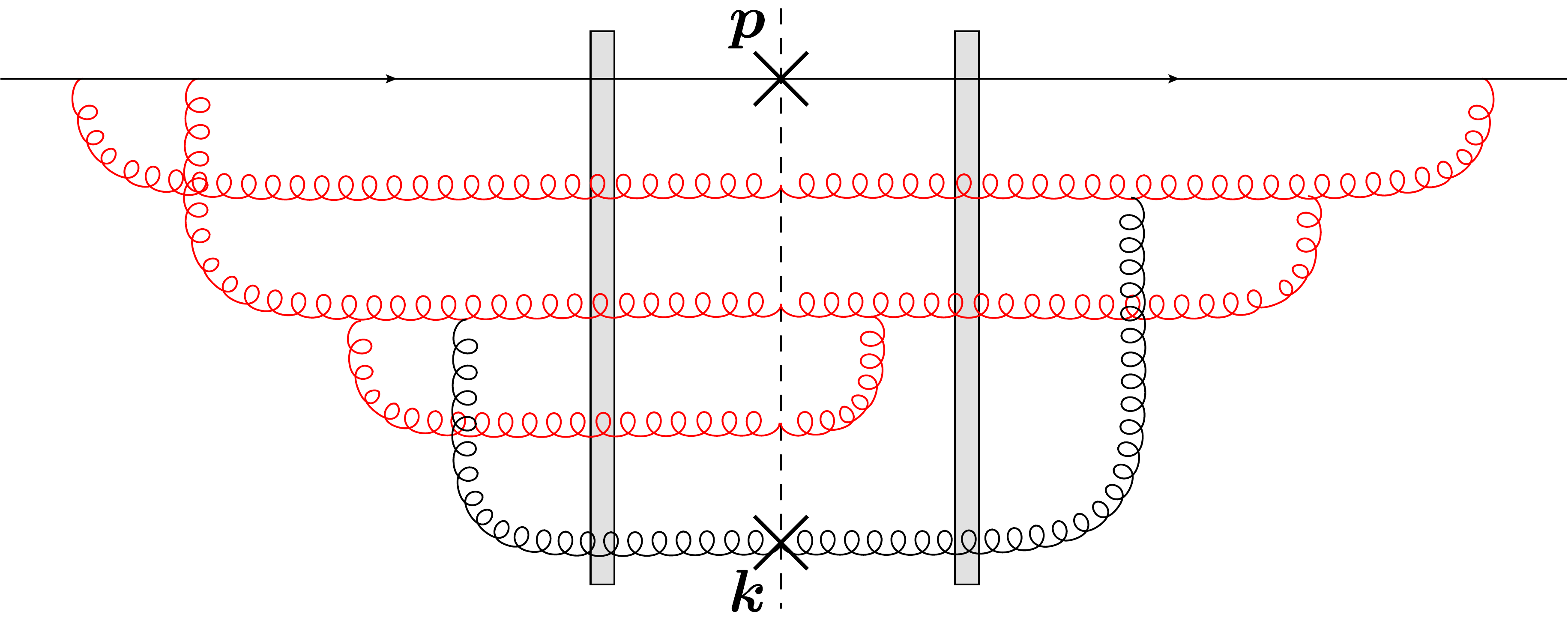}
\qquad\qquad
\includegraphics[width=0.35\textwidth,angle=0]{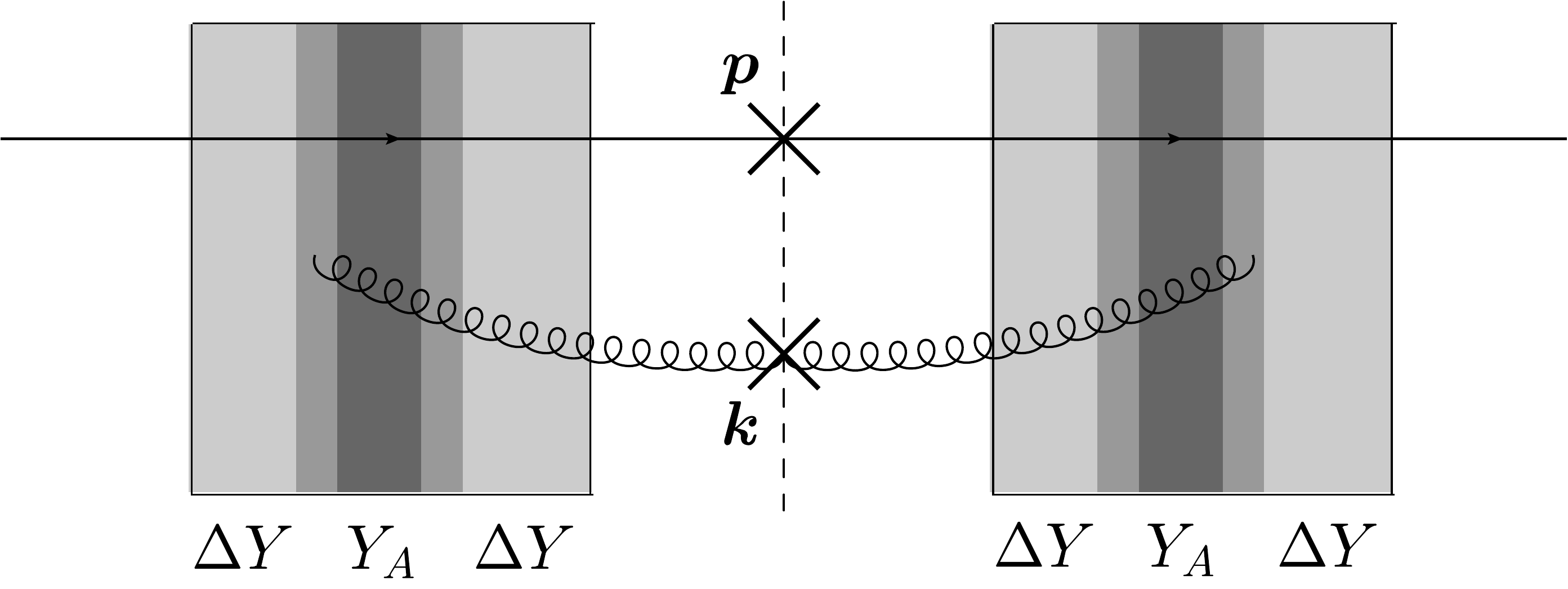}
\end{center}
\vspace*{-0.5cm}
\caption{\label{fig:qg_evol}Evolution at intermediate rapidities between the produced particles for (a) the projectile and (b) the target.}
\vspace*{-0.2cm}
\end{figure}

Fig.~\ref{fig:qg_evol}.a illustrates the viewpoint of projectile evolution which
holds in a frame where the soft produced gluon is relatively slow, the fast quark is a right mover
with rapidity $\Delta Y$ and the nuclear target is a left mover with negative rapidity $-Y_A$,
with $Y_A\equiv Y-\Delta Y$. The soft gluon is emitted by either the quark or any of the gluons within the interval $\Delta Y$. All these produced and unresolved partons can scatter
off the strong target color field. Accordingly, the evolution within $\Delta Y$ cannot be `factorized'
from the collision --- i.e.~it cannot be viewed as a part of the quark wavefunction prior
to scattering. (Factorization is recovered only for a dilute target where one neglects multiple scattering.)

Fig.~\ref{fig:qg_evol}.b illustrates the viewpoint of target evolution which holds in a 
frame where the quark is relatively slow and the target has (negative) rapidity $-Y$. Now all the gluons are part of the target wavefunction, i.e.~they are left-movers. The evolution gluons within $\Delta Y$ are particularly slow, carrying very small longitudinal momenta $k^-$, they are strongly delocalized in $x^+$ and the target looks thicker to the projectile. The measured gluon is only moderately
slow, i.e., it carries a larger value $k^-$, hence it is emitted inside the target at some coordinate $x^+$ (either negative, or positive) of the order of its longitudinal wavelength $1/k^-$. This coordinate is related to the gluon rapidity $Y_A$ as $|x^+|\sim \rme^{Y_A}$.

The target evolution perspective is more convenient for our purposes. The target field is built in layers of $x^+$, with the
inner ones near $x^+=0$ representing the fast and Lorentz contracted valence d.o.f.~and the outermost ones at large $|x^+|$ corresponding to the `wee' gluons with the smallest values
of $k^-$. One evolution step consists in the emission of a gluon which is softer in $k^-$ than all of its ancestors. This adds two new layers to the field at larger values of $|x^+|$, symmetrically located around $x^+=0$. The new fields are {\em random} due to the quantum nature of the gluon emissions. Thus the evolution is naturally stochastic and can be given as a Langevin equation in the space of Wilson lines \cite{Blaizot:2002xy}. 

In this Langevin process we discretize the interval in $Y$ according to $Y=N \epsilon$ and the JIMWLK evolution is equal to a simultaneous left and right rotation of the Wilson lines leading to the recurrence formula (e.g.~for a quark projectile)
 \beq
 \label{un}
 V^{\dagger}_{n,\bx}
 =\rme^{\rmi \epsilon g \alpha^L_{n,\bx}}\,
 V^{\dagger}_{n-1,\bx}\,
 \rme^{-\rmi \epsilon g \alpha^R_{n,\bx}},
 \qquad
 \alpha^L_{n,\bx} = 
 \frac{1}{\sqrt{4\pi^3}}
 \int \dif^2\bz\,  \mcal{K}^i_{\bx\bz} \,
 \nu^{ia}_{n,\bz}\, t^a,
 \qquad
 \alpha^R_{n,\bx} = 
 \frac{1}{\sqrt{4\pi^3}}
 \int \dif^2\bz\,   \mcal{K}^i_{\bx\bz} \,
 U^{ab}_{n-1,\bz} 
 \nu^{ib}_{n,\bz}\, t^a.
 \eeq
The noise $\nu^{ia}_{n,\bz}$ accounts for the charge density and the polarization of the gluons radiated in the evolution step, and which act as sources for $\alpha^{R}_{n}$ and $\alpha^{L}_{n}$. It is a Gaussian white noise local in rapidity, color, spin and transverse coordinates:
 \beq
 \label{nu}
 \big\langle
 \nu^{ia}_{m,\bx} \nu^{jb}_{n,\by}
 \big\rangle
 = \frac{1}{\epsilon}\,
 \delta^{ij} \delta^{ab} 
 \delta_{mn} \delta_{\bx\by}\,.
 \eeq
These noise sources are left movers slower than those produced in the previous steps. Accordingly, the field $\alpha^{R}_{n}$ radiated at negative $x^+$, meaning
{\em ahead} of the shockwave, 
can be caught by the latter and suffer a color-rotation. This is the origin of the adjoint 
Wilson line $U_{n-1}$ in the r.h.s.~of the above equation for $\alpha^{R}_{n}$, which in turn is responsible for generating the BFKL cascade via iterations. The {\em physical} dipole $S$--matrix at $Y$ is finally obtained as
 \beq
 \label{savelang}
 \big\langle
 S_{\bx\by}
 \big\rangle_Y 
 = (1/N_c)\,
 \Big\langle 
 \rmtr \Big\{V^\pd_{N,\by}
 V^{\dagger}_{N,\bx} \Big\}
 \Big\rangle_{\nu}\,,
 \eeq
where the brackets refer to the average over the noise at the intermediate steps $n\le N$. This stochastic procedure, which is equivalent to the 
 CGC average in \eqn{save} and also to solving the B--JIMWLK hierarchy \cite{Gelis:2010nm},
has the advantage to be well suited for numerical implementations \cite{Rummukainen:2003ns,Dumitru:2011vk}. Alternatively, one can rely on Mean Field Approximations \cite{Kovchegov:2008mk,Dumitru:2011vk,Iancu:2011ns,Iancu:2011nj,Alvioli:2012ba}.

The new feature in quark--gluon production is the need to single out from the nuclear wavefunction the gluon with rapidity $Y_A < Y$ which is produced in the final state. One distinguishes between the target evolution up to $Y_A$ and that from $Y_A$ up to $Y=Y_A+\Delta Y$ and then the expectation value entering the cross-section in \eqn{samey} gets replaced by
 \beq
 \label{differenty}
 \int \mcal{D}V_A\,W_{Y_A}[V_A]\,
 H_{\rm prod}^A(\bk) \big\langle 
 S_{\bx\bbx}^<  \big\rangle_{\Delta Y}
 \big|_{\bar{V}_A=V_A}\,.
 \eeq  
$W_{Y_A}[V_A]$ is the target CGC weight function at rapidity $Y_A$ and  $H_{\rm prod}^A(\bk)$ produces the soft gluon at that rapidity; it is obtained from \eqn{hprod}
by replacing $U^\dagger\to U_A^\dagger$, $R^a\to R^a_{A}$ etc.~The quark generating functional
$\big\langle S_{\bx\bbx}^< \big\rangle_{\Delta Y}[V_A, \bar{V}_A]$ for emitting a gluon separated by $\Delta Y$ from the quark can be computed via a Langevin procedure starting at  $\Delta Y=0$ with the initial condition $S^<_{\bx\bbx} [V_A, \bar{V}_A]$, cf. \eqn{s12}.
Specifically, with $\Delta Y=N \epsilon$ and the initial condition $V^{\dagger}_{0,\bx}=V^{\dagger}_{A,\bx}$, one has
 \beq
 \label{S12ev}
 \big\langle {S}^<_{\bx\bbx} 
 \big\rangle_{\Delta Y} [V_A, \bar{V}_A]\,= 
 (1/N_c)\,
 \Big \langle \rmtr\Big\{\bar{V}^\pd_{N,\bbx}
 V^{\dagger}_{N,\bx}\Big\}
 \Big\rangle_{\nu}\,,
 \eeq 
where $V^{\dagger}_{N,\bx}$ is built as shown in \eqn{un}. $\bar{V}^\pd_{N,\bbx}$ is built via
 a similar procedure where all the quantities are `barred' (but such that the noise term
 is the {\em same} in the DA and the CCA: $\bar\nu^{ia}_{n,\bz}=\nu^{ia}_{n,\bz}$  \cite{Iancu:2013uva}).
 
A numerical calculation based on Eqs.~\eqref{differenty}--\eqref{S12ev} is not possible due to the functional initial conditions, but this problem can be circumvented \cite{Iancu:2013uva}. The action of $H_{\rm prod}^A$ on the generating functional involves the sum of four terms like 
 \beq
 \label{rrons}
 R^a_{A,\bu}\, \bar{R}^b_{A,\bbu}\, 
 \big\langle S_{\bx\bbx}^<
 \big\rangle_{\Delta Y}
 \big|_{\bar{V}_A=V_A}
 = (1/N_c)\,
 \Big\langle 
 \rmtr \Big\{
 \big(R^b_{A,\bbu} V_{N,\bbx}^\pd\big) 
 \big(R^a_{A,\bu} V^{\dagger}_{N,\bx}
 \big)\Big\}
 \Big\rangle_{\nu},
 \eeq
and the other terms are obtained from the above using 
$L^a_{A,\bu}= U^{\dagger ab}_{A,\bu}R^b_{A,\bu}$. The dependence of the evolved Wilson lines
$V^{\dagger}_{N} $
and $\bar{V}_{N}$ upon their respective initial conditions $V_A^{\dagger}$ and $ \bar{V}_A$
is generally  complicated, because of the non-linear evolution of the gluons within $\Delta Y$, as reflected by the dependence of the `right' field $\alpha^{R}_{n}$ in \eqn{un} upon $U_{n-1}$ and hence (going backwards along the iterations) upon $U_A$. For illustration consider the one step action of $R^a_{A}$  which gives
 \beq
 \label{Run}
 R^a_{A,\bu} V^{\dagger}_{n,\bx}
 =\rme^{\rmi \epsilon g \alpha^L_{n,\bx}}\,
 \Big(R^a_{A,\bu} V^{\dagger}_{n-1,\bx}\Big)\,
 \rme^{-\rmi \epsilon g \alpha^R_{n,\bx}}
 \,-\,\rmi \epsilon g\,
 \rme^{\rmi \epsilon g \alpha^L_{n,\bx}}\,
 V^{\dagger}_{n-1,\bx}\,
 \Big(R^a_{A,\bu}\,\alpha^R_{n,\bx}\Big)\,.
 \eeq
Within the second term we were allowed to expand the exponential to linear order and the action of $R^a_{A,\bu}$ on $\alpha^R_{n,\bx}$ is an action on the Wilson line $U_{n-1}$, cf. \eqn{un}. This suggests the new strategy: it looks natural to consider a {\em purely numerical} process for both ${V}^{\dagger}_{n,\bx}$ and the {\em bi-local} (in transverse coordinates) quantity $R^a_{A,\bu} V^{\dagger}_{n,\bx}$. The Langevin equation for the ${V}^{\dagger}_{n,\bx}$
is \eqn{un} but extended to the rapidity interval $Y$. (In particular  
$V_A^{\dagger}$ is now built numerically, via 
the stochastic evolution up to the intermediate rapidity $Y_A$.) That for the
bi-local quantity $R^a_{A,\bu} V^{\dagger}_{n,\bx}$ applies to the interval $\Delta Y$ alone. It is conveniently written as a recurrence formula for the {\em color charge
density},
 \beq
 \label{mcalr}
 \mcal{R}^{a}_{n,\bu\bx} 
 \equiv -\rmi V^{\pd}_{n,\bx}
 R^a_{A,\bu} V^{\dagger}_{n,\bx},
 \eeq 
which is a member of the Lie algebra (the subscript $A$ is left implicit, to simplify writing). 
One finds
 \beq
 \label{rlang}
 \mcal{R}^a_{n,\bu\bx}
 =\rme^{\rmi \epsilon g \alpha^R_{n,\bx}}
 \mcal{R}^a_{n-1,\bu\bx}
 \rme^{-\rmi \epsilon g \alpha^R_{n,\bx}}
 -\frac{\rmi \epsilon g}{\sqrt{4\pi^3}}\,
 \rme^{\rmi \epsilon g \alpha^R_{n,\bx}}
 \int \dif^2 \bz\,
 \mcal{K}^i_{\bx\bz}
 U^{bc}_{n-1,\bz} \nu^{ic}_{n,\bz}
 \big[t^b, \mcal{R}^a_{n-1,\bu\bz} \big]\,.
 \eeq   
The initial condition for \eqn{mcalr} is now merely given by $\mcal{R}^a_{0,\bu\bx} = g \delta_{\bu\bx} t^a$. This is local in the transverse plane, but such a property is immediately lost after the first step, as evident in \eqn{rlang}.
This first step also involves $U_0\equiv U_A$, whereas those with $n>1$ will involve the adjoint
Wilson line $U_{n-1}$ built via the parallel process. This is mathematically well defined, and the only numerical obstacle may be the bi-locality of the color charge density.

To the order of accuracy we can expand \eqn{rlang} to order $\epsilon$. Keeping in mind that $\nu \sim 1/\sqrt{\epsilon}$ and averaging over quadratic in $\nu$ terms, we find that local and non-local terms (in the transverse plane) combine to give
 \beq
 \label{rlang2}
 \mcal{R}^a_{n,\bu\bx}
 = \mcal{R}^a_{n-1,\bu\bx}
 +\frac{\rmi \epsilon g}{\sqrt{4\pi^3}}\,
 \int \dif^2 \bz\, 
 \mcal{K}^i_{\bx\bz}
 U^{bc}_{n-1,\bz} \nu^{ic}_{n,\bz}
 \big[t^b, \mcal{R}^a_{n-1,\bu\bx}-
 \mcal{R}^a_{n-1,\bu\bz} \big]
 -\frac{\epsilon g^2 N_c}{8 \pi^3}
 \int \dif^2 \bz\, 
 \mcal{K}^i_{\bx\bz} \mcal{K}^i_{\bx\bz}
 \big(\mcal{R}^a_{n-1,\bu\bx}-
 \mcal{R}^a_{n-1,\bu\bz} \big),
 \eeq 
valid for an arbitrary representation of the color charge density $\mcal{R}^a_{n,\bu\bx}$. In general, the presence of $U$ signals the breaking of $k_{\perp}$--factorization. \eqn{rlang2} simplifies in the limit where there is no scattering: setting $U=1$  leads to the BFKL evolution for the color charge density (unintegrated gluon p.d.f.) and its correlations in the projectile wavefunction. In that limit, the correspondingly simplified Langevin gives the finite-$N_c$ generalization of the color dipole picture  \cite{Mueller:1993rr}. 

\vspace*{-0.2cm}
\providecommand{\href}[2]{#2}\begingroup\raggedright\endgroup








\end{document}